\documentclass[aps,pre,twocolumn,groupedaddress,showpacs,floatfix]{revtex4}
\usepackage[dvips]{graphicx}
\begin{document}
\title{Size scaling of static friction}
\author{O.M. Braun}
\email[E-mail: ]{obraun.gm@gmail.com}
\homepage[Web: ]{http://www.iop.kiev.ua/~obraun}
\affiliation{Institute of Physics, National Academy of Sciences of Ukraine,
  46 Science Avenue, 03028 Kiev, Ukraine}
\author{Nicola Manini}
\affiliation{Dipartimento di Fisica, Universit\`a di Milano,
Via Celoria 16, 20133 Milano, Italy}
\affiliation{International School for Advanced Studies (SISSA),
  Via Bonomea 265, I-34136 Trieste, Italy}
\author{Erio Tosatti}
\affiliation{International School for Advanced Studies (SISSA),
  Via Bonomea 265, I-34136 Trieste, Italy}
\affiliation{CNR-IOM Democritos National Simulation Center,
  Via Bonomea 265, I-34136 Trieste, Italy}
\affiliation{International Centre for Theoretical Physics (ICTP),
P.O. Box 586, I-34014 Trieste, Italy}

\date{\today}
\begin{abstract}

Sliding friction across a thin soft lubricant film typically occurs by
stick-slip, the lubricant fully solidifying at stick, yielding and flowing
at slip.
The static friction force per unit area preceding slip is known from
molecular dynamics (MD) simulations to decrease with increasing contact
area.
That makes the large-size fate of stick-slip unclear and unknown; its
possible vanishing is important as it would herald smooth sliding with a
dramatic drop of kinetic friction at large size.
Here we formulate a scaling law of the static friction force, which for a
soft lubricant is predicted to decrease as $f_m + \Delta f /A^\gamma$ for
increasing contact area $A$, with $\gamma>0$.
Our main finding is that the value of $f_m$, controlling the survival of
stick-slip at large size, can be evaluated by simulations of comparably
small size.
MD simulations of soft lubricant sliding are presented, which verify
this theory.
\end{abstract}
\maketitle

Boundary friction of sliding crystal surfaces 
across atomically thin solid or nearly solid lubricant
layers, of considerable conceptual and 
practical importance, also constitutes an open
physics problem, because the uncertain occurrence of stick-slip makes the
prediction of the overall frictional regime -- stick-slip or smooth
sliding -- rather uncertain \cite{Pbook,BN2006}.
While for hard solid lubricants the answer is known, namely stick-slip for
commensurate and crystallographically aligned interfaces or smooth sliding
for lattice mismatched/misaligned interfaces \cite{note0}, it is not so for
soft solid lubricants.
The latter, with shearing occuring inside the lubricant rather than at the
surface/lubricant interface, represent the commonest case, realized at room
temperature in e.g.\ commercial machine oils confined in between metallic
surfaces.
The possibility 
of smooth sliding would be especially relevant, because of the
accompanying large drop of kinetic friction, often a very desirable
outcome.
The crucial controlling quantity is the magnitude of static friction $f_s$
-- the maximum pulling force reached before slip.
So long as $f_s$ is finite there will be stick-slip;
when $f_s$ drops to zero, there can only be smooth sliding.
Realistic molecular dynamics (MD) simulations of lubricants confined
between atomically flat surfaces generally indicate that stick-slip
prevails for soft lubricants, with consequently high kinetic friction.
However, while in smooth sliding the kinetic friction per unit area is
essentially size independent, its static counterpart $f_s$ may decrease
with increasing contact area $A$ \cite{Pbook,BN2006}.
Despite the increased computer power, the simulated system sizes
\cite{Jagla02,Luan04,Luan05,Vanossi13} are still far too small to establish
conclusively whether in the limit of mesoscopically large size the static
friction will remain finite, and stick-slip will survive with large kinetic
friction, or if it will vanish so that smooth sliding and low dynamic
friction will eventually prevail.
The time-honored approach borrowed from equilibrium statistical mechanics
to this type of question is finite-size scaling \cite{binder_somebook}.
One can for example double repeatedly the size of the simulation cell and
compare the change in the results with some analytically predicted size
dependence from theory.
Given a good scaling prediction, a few simulated sizes are often sufficient
to establish the large-size limit with reasonable accuracy, and in
particular whether static friction will drop to zero and stick-slip will
disappear, or not.

In this Letter we solve this question, first by deriving a size scaling law
for static friction, and then showing that it fits realistic MD
simulations yielding a well defined answer.
Our end result is that (i)~the predicted drop of the size-dependent part of
the static friction per atom $f_s$ is inversely proportional to the linear
size of the contact (i.e.\ to $A^{1/2}$), but that
(ii)~its predicted large-size limit is nonzero, so that stick-slip
will generally survive in soft solid boundary lubrication.

\smallskip
\textit{Scaling theory}.
To start off the theory, we inspect first the dynamics of previous MD
simulations of sliding over soft solid
lubricants~\cite{BN2006,BP2001,BraunManini11}.
These simulations indicate, very reasonably, that unlike hard lubricants
where sliding occurs at the interfaces, plastic motion within the soft
lubricant nucleates typically at some weak point well inside the lubricant
film, such as a point defect, a dislocation, a local incommensurability,
etc.\ (similarly to the ``weakest-link hypothesis'' of fracture mechanics,
e.g., see Refs.~\cite{George52,MSNASZ2012} and references therein).
The static friction force $f_s$ (per substrate atom, i.e.\ $f_s = \sigma_s
L_x L_y/N_s$, where $\sigma_s$ is the shear stress, $L_x$ and $L_y$ are the
sides of the rectangular simulation cell, and $N_s$ is the number of
substrate surface atoms) depends on the given initial (frozen)
configuration.
For a given size $A = L_x \times L_y$ of the simulation cell, different
realizations of the initial configuration will give different $f_s$ values
$f_{s\,1}<f_{s\,2}<...$, where a given value $f_{s\,i}$ is realized with
probability $p_i^{(A)}$.
Now suppose we double the simulation cell.
Slip-motion will again start at the weakest point wherever it is, in either
half of the doubled cell.
Assuming that the new (larger) contact does not develop new thresholds, the
probability that the doubled cell fails at threshold $f_{s\,i}$ equals the
sum of the probability that failure occurs precisely at this threshold
$f_{s\,i}$ in both halfs plus the probability that in one half the
threshold is $f_{s\,i}$ and in the other half it is some larger
$f_{s\,j}$:
\begin{equation}
\label{eq01}
p_i^{(2A)} = \left( p_i^{(A)} \right)^2
+ 2 p_i^{(A)} \sum_{j>i} p_j^{(A)}.
\end{equation}
The factor 2 accounts for the the two symmetric realizations of $f_{s\,j}$
and $f_{s\,i}$ in the two halves.

By iteration of Eq.~(\ref{eq01}), we can find the probability
$p_i^{(\Lambda A)}$ for larger and larger cell size $\Lambda A \equiv 2^n
A$, with $n=0,1,\ldots$
Given the resulting distribution, one can calculate the {\em average}
static threshold for the $\Lambda A$ cell by
\begin{equation}
\label{eq02}
\bar f_s (\Lambda) = \sum_i p_i^{(\Lambda A)} f_{si} \,.
\end{equation}

To illustrate this approach, consider the simple instructive example where
only two thresholds $f_{s\,1} = f_m$ and $f_{s\,2} = f_m + \Delta f > f_m$
occur, with probabilities $p_1$ and $p_2$.
For the doubled cell, we have four possible thresholds realizations:
$(f_{s\,1},f_{s\,1})$ with probability $p_1^2$,
$(f_{s\,1},f_{s\,2})$ with probability $p_1 p_2$,
$(f_{s\,2},f_{s\,1})$ with probability $p_2 p_1$, and
$(f_{s\,2},f_{s\,2})$ with probability $p_2^2$.
Accordingly, the doubled cell fails at the lower threshold with probability
$p_1^2 + 2p_1 p_2$, and at the upper threshold with probability $p_2^2$.
Indicate with $p_1^{(A)} = 1 - \alpha$ and $p_2^{(A)} = \alpha <1$.
The iteration chain is $p_2^{(2 \Lambda A)} = \left[ p_2^{(\Lambda A)}
  \right]^2$, with solution $p_2^{(\Lambda A)} = \alpha^{\Lambda}$, and
thus $p_1^{(\Lambda A)} = 1 - \alpha^{\Lambda}$.
Accordingly, the average static friction approaches the minimum threshold
$f_m$ exponentially in $\Lambda$:
\begin{equation}
\label{eq03}
\bar f_s (\Lambda) - f_m = \alpha^{\Lambda} \Delta f
= e^{\Lambda \ln \alpha} \Delta f \,.
\end{equation}

\begin{figure}
\includegraphics[clip,width=8cm]{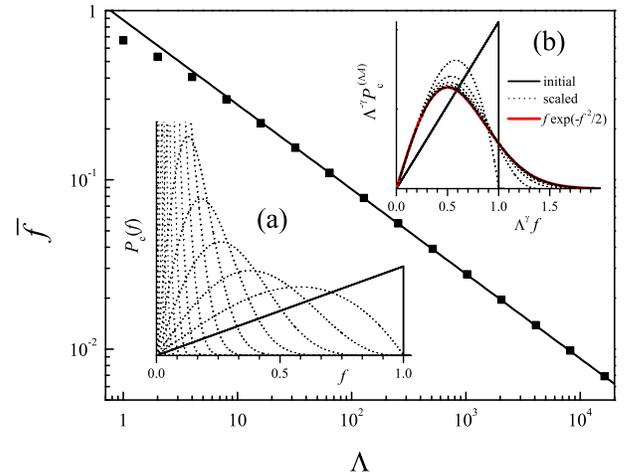}
\caption{\label{B02}(color online)
Iteration of Eq.~(\ref{eq04}) starting from a sawtooth initial distribution
$P_c^{(A)} (f) = 2 \, f \, \theta(f) \,\theta(1-f)$, where $\theta(x)$ is
the Heaviside step function.
Inset (a) shows successive iterations of $P_c^{(\Lambda A)}(f)$ for
$\Lambda=1,2,4,8,\dots$
Inset (b) displays the scaled distribution $\Lambda^{-\gamma}
P_c^{(\Lambda A)}(\Lambda^{-\gamma} f)$ compared to the infinite-size
solution (\ref{gfunction}).
The main graph shows the average static friction $\bar f$ excess (above the
minimum $f_m$) as a function of contact size.
}
\end{figure}

When we replace the discrete thresholds $f_{s\,i}$ with a more realistic
continuous distribution with probability $P_c^{(A)}(f_s)$, the first,
quadratic, contribution in Eq.~(\ref{eq01}) can be neglected, and the
iteration equation takes the form
\begin{equation}
\label{eq04}
P_c^{(2A)}(f_s) = 2 P_c^{(A)}(f_s) \times
\int_{f_s}^{\infty} P_c^{(A)}(f_s') \, df_s'
\,.
\end{equation}
Figure~\ref{B02}(a) illustrates an example of the numerical iteration of this
equation.
Simulations suggest that for large $\Lambda$ the distribution
$P_c^{(\Lambda A)}(f_s)$ tends to approach some universal shape, with
little dependency on the small-size distribution $P_c^{(A)}(f_s)$, once it
is rescaled appropriately.
A similar scaling behavior was proven for the strain distributions of the
(related) fiber-bundle models
\cite{Newman91,Smalley85,Newman94,Newman01,note1}.
In these 
models the conditions for the emergence of a critical
point, i.e.\ a finite stress in the large-scale limit, were investigated
under the assumption of a nonzero single-fiber breaking probability for
arbitrarily small stress.
Here instead we consider that more generally the minimal contact (``single
fiber'') distribution of unpinning forces $f_s$ can start off at a minimum
$f_m$ which can be nonzero.
Iteration of Eq.~(\ref{eq04}) guarantees that for any contact size $\Lambda
A$ the distribution $P_c^{(\Lambda A)}(f_s)$ vanishes below the same $f_m$ as
$P_c^{(A)}(f_s)$: scaling preserves $f_m$.
To address the scaling of the distribution above $f_m$, it is convenient to
introduce $f=f_s-f_m$.
Let us assume that at large $\Lambda$ the 
normalized probability distribution scales as
\begin{equation}
\label{eq05}
P_c^{(2\Lambda A)}(f) = a P_c^{(\Lambda A)}(a f) \,,
\end{equation}
where $a>1$ is a constant.
By substituting Eq.~(\ref{eq05}) into Eq.~(\ref{eq04}), for the large-size
distribution $g(f) = \lim_{\Lambda \to \infty} \Lambda^{-\gamma}
P_c^{(\Lambda A)}\left(\Lambda^{-\gamma}\,f\right)$,
with $\gamma=\log_2 a$, we obtain the following equation:
\begin{equation}
\label{eq06}
a \, g(af) = 2 g(f) \int_{f}^{\infty} g(f') \, df' ,
\end{equation}
or 
\begin{equation} 
\label{eq07}
 \frac{a \, g(af)}{g(f)} = 2\int_f^{\infty} g(f') \, df'.
\end{equation}
Differentiating both sides with respect to $f$, we get
\begin{equation}
\label{eq08}
a^2 g' (af) \, g(f) - a \, g(af) \, g'(f) +2 g^3 (f) =0
\,.
\end{equation}
The solutions of this equation depend on a single feature of the
distribution $g(f)$, namely its small-$f$ behavior.
More precisely, assuming that $g(f)=\sum_{k=k_0} c_k f^k$, with $k_0>-1$,
we have that
\begin{eqnarray}\label{scalingpower}
a &=& 2^\gamma \qquad {\rm with}\ \gamma = (1+k_0)^{-1} \,,
\\\label{gfunction}
g(f)&=& c_{k_0} f^{k_0} \exp\left(-\frac{c_{k_0}  f^{1+k_0}}{1+k_0}\right)
\end{eqnarray}
solve Eq.~(\ref{eq08}).
Figure~\ref{B02}(b) demonstrates the approach of the scaled distributions
to the function $g(f) = f \, \exp (- f^2 /2)$ (with $a = \sqrt{2}$)
obtained by starting off with an initial distribution $P_c^{(A)}(f) = 2 \,
f \, \theta(f) \, \theta(1-f)$, i.e.\ with $k_0=1$, $c_1=2$.

The scaling theory makes the following predictions:
(i)~as scaling preserves $f_m$, it is possible to predict the minimum
threshold $f_m$ from an evaluation of $P_c^{(A)}(f)$ at the smallest
contact size;
(ii)~the iteration defined by Eq.~(\ref{eq04}) preserves the leading term in
the $f$ power expansion of $P_c^{(A)}(f)$ above $f_m$;
(iii)~regardless of the overall shape of the small-size threshold
distribution, for large size the distribution acquires the ``universal''
shape of Eq.~(\ref{gfunction});
(iv)~its width $\Delta f^{(\Lambda A)}$ scales down as an inverse power law
of $\Lambda$;
(v)~this power law is dictated uniquely by the leading power law with which
the arbitrary-size threshold distribution behaves for $f_s$ near $f_m$;
(v)~as $\Lambda$ increases, the average friction force $\bar f_s$
approaches $f_m$ according to the law
\begin{equation}
\label{scaling_law}
\bar f_s(\Lambda) - f_m \simeq
\left[\bar f_s(\Lambda=1)-f_m\right]
\Lambda^{-\gamma}
\,.
\end{equation}
In the example of Fig.~\ref{B02}, this relation yields a mean excess static
friction scaling as the inverse square root of size $\bar f_s -f_m \propto
\Lambda^{-1/2}$.

\begin{figure}
\includegraphics[clip,width=8cm]{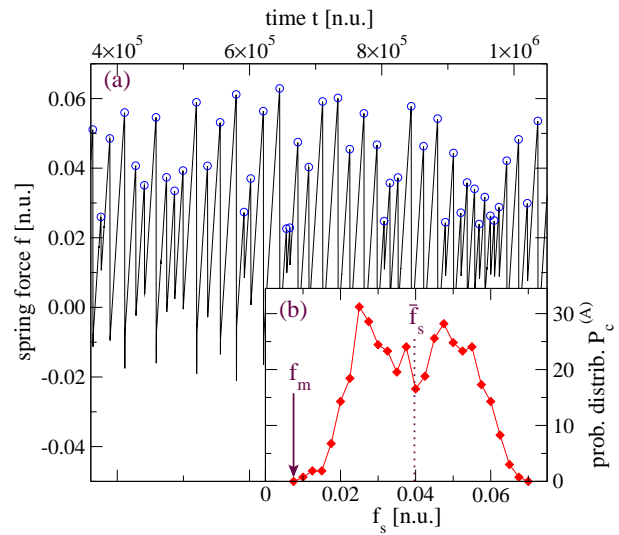}
\caption{\label{SSS}(color online)
(a) The time evolution of the spring force during a segment of the
  simulated stick-slip dynamics of the $12 \times 11$ substrate model
  driven at speed 0.01~n.u., with an applied load of 0.1~n.u.\ per
  rigid substrate atom; circles mark the stick-to-slip transitions where
  the individual static-friction thresholds $f_{si}$ are extracted.
(b) The probability distribution of the static-friction thresholds
  $P_c^{(A)} (f_s)$ as estimated by a histogram of the $f_{si}$ values.
  A dotted line marks the mean value $\bar f_s$, and an arrow marks the
  estimated $f_m$.
}
\end{figure}

\smallskip
\textit{Simulation}.
To validate our prediction with MD simulations, we use our previously
developed model \cite{BN2006,BP2001,BraunManini11}.
Each of the two substrates is modeled by two atomic layers, one rigid and
one deformable.
In the minimum size simulation of contact area $A$, these substrates are
composed by $12 \times 11$ atoms arranged in a square lattice.
The space between the substrates is filled by three incomplete layers of
lubricant atoms
(to prevent crystallization of the lubricant, we put approximately 90\% of
the atoms which would complete 3 perfect monolayers).
All atoms interact according to the Lennard-Jones (LJ) potential.
The strength of the lubricant-lubricant interaction is $V_{ll}=1/9$ (in
dimensionless ``natural'' units, n.u., defined for example in
Refs.~\cite{BP2001,BN2006}), while the lubricant-substrate interaction is
much stronger, $V_{sl}=1/3$.
The equilibrium distance of the LJ lubricant-lubricant interatomic
potential is
$r_{ll}/a_s = 3.95/3$
(i.e., the solid lubricant is incommensurate with the substrate).
These parameters correspond  to a soft lubricant.
Once the thin soft lubricant film is interposed between the sliders,
sliding takes place inside the lubricant (as opposed to hard lubricants
where the sliding would take place at the interfaces) and the film melts
during sliding, realizing the melting-freezing mechanism of stick-slip
motion.
The bottom substrate is kept fixed, while the rigid top slider layer is
pressed with a load of $0.1$~n.u.\ per substrate atom (representing a
pressure in the order $100$~MPa
if the model represented a noble gas solid lubricant between metal
surfaces) and driven through a spring of elastic constant $k = 3 \times
10^{-4}$~n.u.\ per atom with a velocity $v$.
We carry out simulations at driving velocities $v = 3\times 10^{-3} \div
3\times 10^{-2}$~n.u.
These velocities are sufficiently small that the system exhibits stick-slip
motion, as illustrated in Fig.~\ref{SSS}(a).
We carry out runs of duration exceeding $10^7$~n.u.\ for the smallest-size
system ($12 \times 11$), representing $\Lambda =1$.
By extracting the ``static friction'' thresholds $f_{si}$ marked by circles
in Fig.~\ref{SSS}(a) in correspondence to the peaks in $f(t)$ preceding
each slip event, we obtain a set $\{ f_{si} \}$ sampling the probability
distribution $P_c^{(A)} (f_s)$.
We evaluate this distribution by means of a histogram of 1063 thresholds
obtained during a long run, and shown in Fig.~\ref{SSS}(b).
Although the detailed behavior near the minimum threshold $f_m$ is
naturally affected by limited statistics, the data are consistent with a
distribution staring off at $f_m\simeq 0.0075$, with an approximately
linear slope ($k_0=1$), which produces an exponent $\gamma=1/2$.
 From the threshold distribution $P_c^{(A)}(f_s)$ we extract the mean value
$\bar f_s = 0.0397$, marked by a dotted line in Fig.~\ref{SSS}(b).

\begin{figure}
\includegraphics[clip,width=8cm]{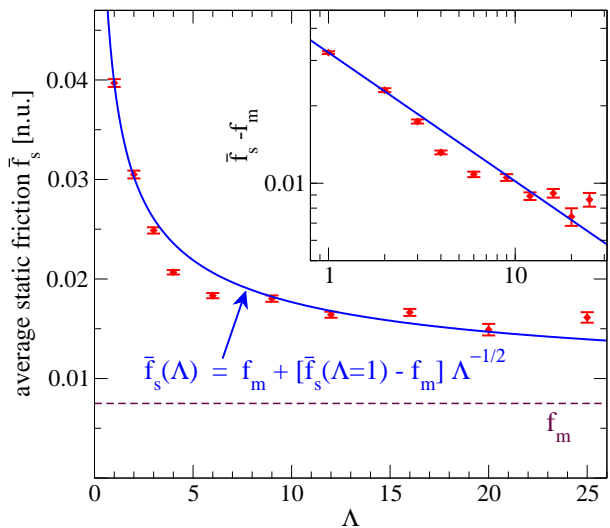}
\caption{\label{size_scaling}(color online)
The average static friction force per substrate atom as a function of the
system size, as obtained from molecular dynamics simulations in the same
conditions as those of Fig.~\ref{SSS}, with several sizes multiple of the
$12 \times 11$ substrate model which is represented by the first $\Lambda
=1$ point in figure.
The solid line shows the scaling law, Eq.~(\ref{scaling_law}).
Inset: log-log scale.
The dashed line marks $f_m$, estimated by the lowest observed slip
threshold in all simulations of all sizes.
}
\end{figure}

Having thus characterized the small-area sliding behavior, we proceed to
increase the area in order to track the size-induced changes.
The cell is successively increased to
$\Lambda = 2 = 2 \times 1$, $3=3 \times 1$, $4=2\times 2$,
$6=3 \times 2$, $9=3\times 3$, $12=4 \times 3$,
$16=4\times 4$, $20=5 \times 4$, and $25=5\times 5$.
The results are presented in Fig.~\ref{size_scaling}.
As expected, the average static friction decreases with system size.
As was noted, it would not be feasible to extract a large-size limit in the
absence of a scaling law.
We find that scaling law~(\ref{scaling_law}) with $\gamma=1/2$ fits
the simulation results with reasonable accuracy.
The static friction tends to a finite large-size value $f_m > 0$, and
therefore stick-slip will survive at macroscopic size.

\smallskip
\textit{Discussion}.
We just arrived at the conclusion that once $f_m$ starts off nonzero, $f_s$
will converge to $f_m > 0$ and static friction will not disappear in the
large-size limit.
One should however not jump to the conclusion that once the
static friction threshold distribution starts from zero, $f_m =0$, the
amplitude of stick-slip jumps of $f(t)$ will drop to zero, and stick-slip
friction will necessarily disappear in the limit of large contact area.
There are two reasons why this is not generally true.
The first reason resides in statics, and follows from elasticity of the
substrate.
The size $\lambda_c$ of a domain that can be considered as rigid and slides
as a whole, is determined by the elastic correlation length
\cite{CN1998,BPST2012}.
%
Therefore, the average static friction force should reach a plateau for
sizes $L \simeq A^{1/2} \agt \lambda_c$ \cite{note2}.
%
The second reason follows from kinetics.
When the sliding motion starts off at some weak contact site, it may either
die off, or spread over the whole interface with some speed $c$.
%
This process takes a finite time $\tau \sim L/c$.
If at a given driving velocity $v$, $\tau$ is of order or larger then the
time between successive slips $\tau_{ss} \sim f_s (L) /(kv)$, then the
local sliding initiated by this weak contact will lose its role and
effectiveness for all sizes $L > \lambda_d$, where $\lambda_d = (c/v) \,
f_s (\lambda_d) /k$, hence the static friction will saturate rather than
decrease further 
to $f_m$ as predicted by the scaling theory Eq.~(\ref{scaling_law}).

This leads us to ask more generally what is lost in the scaling approach?
We assumed that the doubled cell has the same set of thresholds
$\{f_{si}\}$ as the original one.
Yet a larger cell may develop new collective excitations, e.g., a
dislocation loop of a size~$> L$.
However, if the original cell $A$ is large enough, we may still safely
extrapolate its distribution to a mesoscopic size, where the master
equation approach \cite{BP2008} is applicable.

Another situation where 
a different behaviour is expected
is the Aubry incommensurate superlubric state, which has zero static
friction threshold in the infinite system \cite{BKbook}.
That state, corresponding to $f_m =0$ in our theory, 
occurs preferentially for hard lubricants whose interior does
not develop a shear band, and which do not melt during sliding.

Experimentally, the scaling behavior predicted here could be probed by
comparing friction-force microscopy realizations with tips of different
curvature radii \cite{Meyer98,Schwarz97} sliding on a surface covered by a
lubricant close to its melting points, e.g.\ an octamethylcyclotetrasiloxan
or an ionic liquid at room temperature, or a noble-gas layer at a cryogenic
temperature.
The technological bottomline lesson, finally, is that stick-slip could be
attenuated by reducing the smallest threshold force, for example by
promoting extra defects in the lubricant film by additives or other means.

\smallskip
\textit{Acknowledgments}.
We thank A.\ Vanossi and M.\ Urbakh for helpful discussions.
O.M.B.\ was supported in part by grants from the Cariplo Foundation
managed by the Landau Network -- Centro Volta,
whose contribution is gratefully acknowledged.
He also thanks ICTP and SISSA (Trieste), and the
University of Milan, for hospitality during this project.
E.T. acknowledges support by SINERGIA Project CRSII2 136287/1.



\begin{thebibliography}{99}
\bibitem{note2}
%
At distances $r > \lambda_c$, where different regions may undergo different
displacements, the sliding of one domain (``fiber failure'') leads to an
increase of the shear stress on neighboring domains (``fibres'') and may
stimulate their sliding too.
%
This effect may be treated in the framework of the fiber-bundle model
\cite{Newman91,Smalley85,Newman94,Newman01} and should lead to the further
decrease of the static friction to zero, but at an extremely slow rate,
$\bar f_s (\Lambda) \propto 1/ \log (\log \Lambda)$.
\end{thebibliography}
\end{document}